\documentclass[showpacs]{revtex4} 
\usepackage{epsfig,amsmath,amssymb,graphicx,upgreek,textcomp}
\usepackage{natbib}
\usepackage[english]{babel} %
\usepackage{indentfirst}

\begin{document}

\vspace{0mm}
\title{Bose-Einstein condensation at constant pressure} %
\author{Yu.M. Poluektov}
\email{yuripoluektov@kipt.kharkov.ua (y.poluekt52@gmail.com)} %
\affiliation{National Science Center ``Kharkov Institute of Physics and Technology'', 61108 Kharkov, Ukraine} %

\begin{abstract}
In the weakly non-ideal gas model \cite{YP1}, the Bose-Einstein
condensation at constant pressure is considered. The temperature of
transition to the state with condensate is found. Temperature
dependences of the total density and condensate density, the energy,
entropy and heat capacities are calculated.
\newline%
\end{abstract}
\maketitle

\vspace{0mm}
\section{Introduction}\vspace{-0mm} 
When considering an ideal Bose gas of $N$ particles, it is usually
assumed that the volume $V$ is fixed and, consequently, the total
density $n=N/V$ is constant. The Bose-Einstein condensation under
the condition of constant pressure is usually not considered. This
is due to the fact that in the ideal gas model below the
condensation temperature the pressure is determined only by
temperature. In this paper, the condensation at constant pressure is
considered in the weakly non-ideal Bose gas model \cite{YP1}, and
the temperature dependences of the energy, entropy, isochoric and
isobaric heat capacities, the total density and condensate density
are found.

\section{Model of an ideal Bose gas}\vspace{-0mm}
Let us first consider the ideal gas approximation. The thermodynamic
functions of an ideal Bose gas above the condensation temperature
can be expressed through the special Bose functions
\begin{equation} \label{01}
\begin{array}{l}
\displaystyle{%
  {\rm B}_\nu(t)=\frac{1}{\Gamma(\nu)}\int_0^\infty\!\frac{z^{\nu-1}dz}{e^{z-t}-1}=\sum_{n=1}^\infty\frac{e^{nt}}{n^\nu}, %
}%
\end{array}
\end{equation}
where $\Gamma(\nu)$ is the gamma function, $\nu=1/2, 3/2, 5/2$. %
In particular, the particle number density and pressure are given by the formulas %
\begin{equation} \label{02}
\begin{array}{l}
\displaystyle{%
  n=\frac{N}{V}=\frac{g}{\Lambda^3}\,{\rm B}_{3/2}(t), %
}%
\end{array}
\end{equation} \vspace{-04mm}
\begin{equation} \label{03}
\begin{array}{l}
\displaystyle{%
  p=\frac{gT}{\Lambda^3}\,{\rm B}_{5/2}(t), %
}%
\end{array}
\end{equation}
where $g\equiv 2s+1$, $s$ -- the spin of a particle, $T$ -- the
temperature. In calculations, where necessary, we assume $g=1$. The
parameter $t=\mu/T$, where $\mu$ is the chemical potential, can take
values $t\le 0$. Equations (\ref{02}),\,(\ref{03}) include the de
Broglie thermal wavelength of a particle, determined by the formula
\begin{equation} \label{04}
\begin{array}{c}
\displaystyle{\hspace{0mm}%
  \Lambda\equiv\Lambda(T)=\bigg(\frac{2\pi\hbar^2}{mT}\bigg)^{\!\!1\!/2}, %
}%
\end{array}
\end{equation}
where $m$ is the mass of a particle.

If one considers a gas of $N$ particles in a constant volume $V$,
and, consequently, at a constant density $n=N/V$, then, according to
Einstein, when the chemical potential turns to zero, at the temperature %
\begin{equation} \label{05}
\begin{array}{c}
\displaystyle{\hspace{0mm}%
  T_B=\frac{2\pi\hbar^2}{m}\bigg(\frac{n}{g\zeta(3/2)}\bigg)^{\!\!2/3} %
}%
\end{array}
\end{equation}
($\zeta(3/2)$ is the value of the Riemann zeta function) the
particles begin to move to the ground energy level with zero
momentum, so that at $T<T_B$ the total number of particles is equal
to the sum of the particles in the condensate $N_B$ and the
over-condensate particles $N'$: $N=N_B+N'$.

\begin{figure}[h!]
\vspace{-1mm}  \hspace{0mm}
\includegraphics[width = 7.35cm]{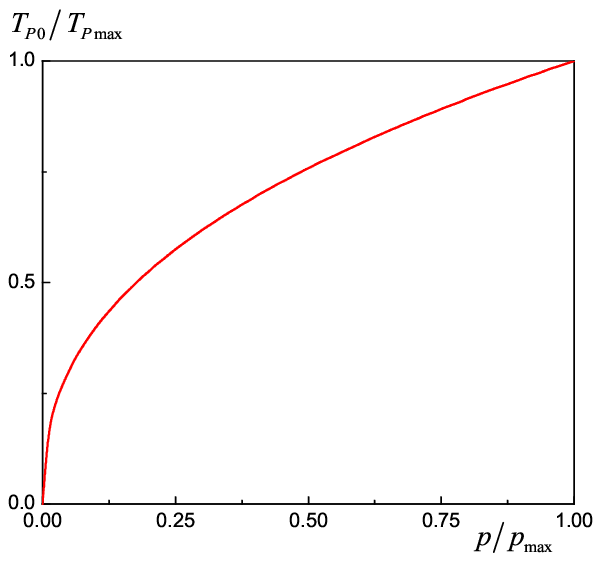} 
\vspace{-4mm} %
\caption{\label{fig01} 
Dependence of the condensation temperature on pressure in the ideal gas model. %
}%
\end{figure}

Experimentally, it is not difficult to realize conditions under
which a gas of $N$ particles will be at a constant pressure $p$. In
this case, the volume and, consequently, the density will change
with temperature. Under such conditions, the chemical potential will
turn to zero at the temperature
\begin{equation} \label{06}
\begin{array}{c}
\displaystyle{\hspace{0mm}%
  T_{P0}=\bigg(\frac{p}{g\zeta(5/2)}\bigg)^{\!\!2/5}\bigg(\frac{2\pi\hbar^2}{m}\bigg)^{\!\!3/5}, %
}%
\end{array}
\end{equation}
where $\zeta(5/2)=1.341$. As is known, at the parameters of $^4$He,
for which the mass of the atom $m=6.648\cdot\!10^{-24}$\,g and the
mass density $\rho=0.145\,\,{\rm g}\!\cdot\!{\rm cm}^{-3}$, the
temperature (\ref{05}) is close to the superfluid transition
temperature $T_B=3.13$\,K. At the maximum pressure $p_{{\rm
max}}=25$\,atm, when
$^4$He remains liquid, the temperature (\ref{06}) $T_{P{\rm max}}=6.07$\,K. %
Figure \ref{fig01} shows the dependence of temperature $T_{P0}$ on
pressure. Below this temperature, the total particle number density
is equal to the sum of the particle number density in the condensate
$n_B$
and the over-condensate density $n'=\zeta(3/2)\big(g/\Lambda^3\big)$: %
\begin{equation} \label{07}
\begin{array}{l}
\displaystyle{%
  n=n_B+\frac{g}{\Lambda^3}\,\zeta(3/2), %
}%
\end{array}
\end{equation}
and the total pressure is equal to the sum of the condensate
pressure $p_B$ and the pressure of over-condensate particles $p'=\zeta(5/2)\big(gT/\Lambda^3\big)$: %
\begin{equation} \label{08}
\begin{array}{l}
\displaystyle{%
  p=p_B+\frac{gT}{\Lambda^3}\,\zeta(5/2). %
}%
\end{array}
\end{equation}
Since the ideal gas approximation is not thermodynamically correct
for a Bose gas with condensate \cite{YP1}, and the pressure in the
state with condensate is determined only by temperature, the
quantities $n_B, p_B$ cannot be calculated in such a model. For
their calculation, the interaction between particles should be taken into account. %

\section{Model of a weakly non-ideal Bose gas}
\begin{figure}[h!]
\vspace{-1mm}  \hspace{0mm}
\includegraphics[width = 7.05cm]{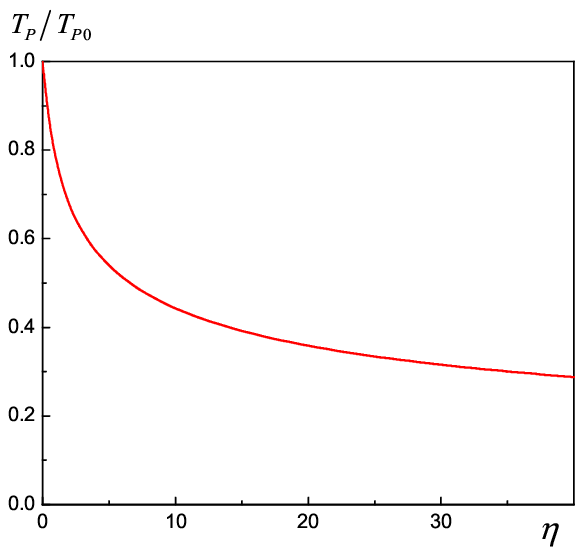} 
\vspace{-4mm} %
\caption{\label{fig02} 
Dependence of the ratio of the condensation temperatures in the case
of account of interaction $T_P$ and in the ideal gas model $T_{P0}$
on the dimensionless parameter $\eta$ (12).
}%
\end{figure}

\vspace{-0mm}
The transition to the weakly non-ideal gas model \cite{YP1} can be
accomplished by replacing the chemical potential with the effective
chemical potential $\mu\rightarrow\mu_*=\mu-2\upsilon n$, where the
interaction constant $\upsilon>0$, so that in the following we
assume $t=\mu_*/T$. In addition, there appears a term in the
formulas for pressure and energy that depends on the total density
\begin{equation} \label{09}
\begin{array}{l}
\displaystyle{%
  p=\upsilon n^2 +\frac{gT}{\Lambda^3}\,{\rm B}_{5/2}(t),  %
}%
\end{array}
\end{equation} \vspace{-01mm}
\begin{equation} \label{10}
\begin{array}{l}
\displaystyle{%
  \frac{E}{N}=n^{-1}\bigg[\upsilon n^2 +\frac{3}{2}\frac{gT}{\Lambda^3}\,{\rm B}_{5/2}(t)\bigg]. %
}%
\end{array}
\end{equation}
When the interparticle interaction is taken into account, the
temperature of Bose-Einstein condensation $T_P$ is also determined
by the condition $t=0$ and is found from the equation
\begin{equation} \label{11}
\begin{array}{l}
\displaystyle{%
  p=\upsilon\frac{g^2}{\Lambda_p^6}\big[\zeta(3/2)\big]^2 + \frac{gT_P}{\Lambda_p^3}\zeta(5/2),  %
}%
\end{array}
\end{equation}
where $\Lambda_p\equiv\Lambda(T_P)$. This equation can be
represented in the form
\begin{equation} \label{12}
\begin{array}{l}
\displaystyle{%
  \eta\,\bigg(\frac{T_P}{T_{P0}}\bigg)^{\!3} + \bigg(\frac{T_P}{T_{P0}}\bigg)^{\!5/2} - 1 = 0,   %
}%
\end{array}
\end{equation}
where the dimensionless parameter is introduced
\begin{equation} \label{13}
\begin{array}{l}
\displaystyle{%
  \eta\equiv \frac{25}{4}\frac{\upsilon p}{T_{P0}^2\sigma_0^2}.  %
}%
\end{array}
\end{equation}
Here and in the following we use the designation $\sigma_0\equiv
\frac{5}{2}\frac{\zeta(5/2)}{\zeta(3/2)}\approx 1.283$. The value of
$\sigma_0$ in its physical meaning is equal to the entropy per one
over-condensate particle. The parameter (\ref{13}) increases
linearly with increasing the interaction constant and slowly
increases with increasing pressure $\eta\sim p^{1\!/5}$. Figure \ref{fig02}\, %
shows the dependence of the temperature ratio $T_P/T_{P0}$ on the parameter $\eta$. %
As we can see, this ratio monotonically decreases with increasing $\eta$. %

\section{Weakly non-ideal Bose gas with condensate at constant pressure}
\begin{figure}[h!]
\vspace{-1mm}  \hspace{0mm}
\includegraphics[width = 7.15cm]{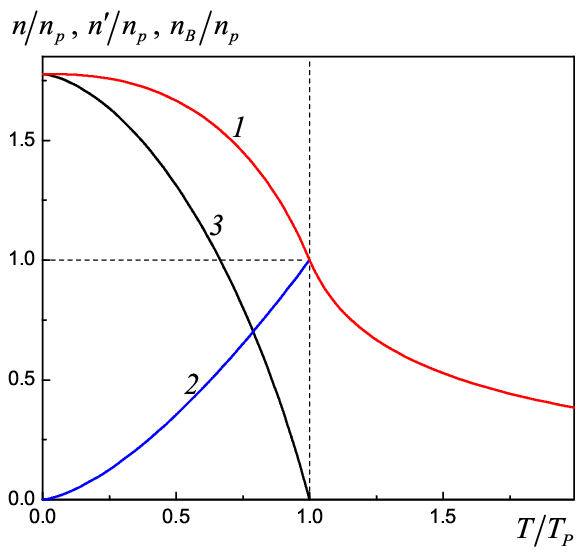} 
\vspace{-4mm} %
\caption{\label{fig03} 
Temperature dependences of {\it 1} -- the total density $n(T)/n_p$, %
{\it 2} -- the over-condensate density $n'(T)/n_p$,  %
{\it 3} -- the condensate density $n_B(T)/n_p$ at $\eta=0.5$. %
}%
\end{figure}

\vspace{-0mm}
Let us consider the state of the Bose gas in the temperature region
$T<T_P$. Here, the temperature dependence of the total density
$n=n(T)$ is determined by the equation
\begin{equation} \label{14}
\begin{array}{l}
\displaystyle{%
  p=\upsilon n^2 +\frac{gT}{\Lambda^3}\,\zeta(5/2), %
}%
\end{array}
\end{equation}
and the density of over-condensate particles by the equation
\begin{equation} \label{15}
\begin{array}{l}
\displaystyle{%
  n'(T)= \frac{g}{\Lambda^3}\,\zeta(3/2), %
}%
\end{array}
\end{equation}
so that the density of particles in the condensate
$n_B(T)=n(T)-n'(T)$. The total density at zero temperature $n(0)$,
which coincides with the condensate density, is determined by the
pressure and the interaction constant $n^2(0)=p/\upsilon$. The
dependence of the total density on temperature can be represented as
\begin{equation} \label{16}
\begin{array}{l}
\displaystyle{%
  n(T)=n(0)\left\{1-\bigg[1-\frac{n_p^2}{n^2(0)}\bigg]\bigg(\frac{T}{T_P}\bigg)^{\!5/2}\right\}^{\!1\!/2}, %
}%
\end{array}
\end{equation}
where
\begin{equation} \label{17}
\begin{array}{l}
\displaystyle{%
  n_p= \frac{g}{\Lambda_p^3}\,\zeta(3/2) %
}%
\end{array}
\end{equation}
is the particle number density at the condensation temperature
$T_P$. The density ratio
\begin{equation} \label{18}
\begin{array}{l}
\displaystyle{%
  \frac{n_p}{n(0)}=\bigg(\frac{T_P}{T_{P0}}\bigg)^{\!3/2}\sqrt{\eta}\,, %
}%
\end{array}
\end{equation}
with account of Eq.\,(\ref{12}), is completely determined by the
value of the parameter $\eta$ (\ref{13}). The temperature dependence
of the density of over-condensate particles has the form
\begin{equation} \label{19}
\begin{array}{l}
\displaystyle{%
  \frac{n'(T)}{n(0)}=\sqrt{\eta}\,\bigg(\frac{T_P}{T_{P0}}\bigg)^{\!3/2}\bigg(\frac{T}{T_{P}}\bigg)^{\!3/2}. %
}%
\end{array}
\end{equation}
The temperature dependences of the densities are shown in Figure \ref{fig03}. %

Let us consider the temperature dependences of entropy and heat
capacities in the state with condensate. The entropy per one
particle is given by the formula %
$S/N=(5/2)\big[g\zeta(5/2)/n\Lambda^3\big]$ \cite{YP1}. %
Taking into account the temperature dependence of the total density
(\ref{16}), we obtain the temperature dependence of entropy at constant pressure %
\begin{equation} \label{20}
\begin{array}{l}
\displaystyle{%
  \frac{S}{N}=\sigma_0\frac{n_p}{n(0)}\bigg(\frac{T}{T_P}\bigg)^{\!3/2}\! %
  \left\{1-\bigg[1-\frac{n_p^2}{n^2(0)}\bigg]\bigg(\frac{T}{T_P}\bigg)^{\!5/2}\right\}^{\!-1\!/2}. %
}%
\end{array}
\end{equation}
To calculate heat capacities, one should take into account that the
differentials of entropy and pressure have the form
\begin{equation} \label{21}
\begin{array}{l}
\displaystyle{%
  dS = S\,\frac{dV}{V} + \frac{3}{2}\,S\,\frac{dT}{T}, %
}%
\end{array}
\end{equation} \vspace{-02mm}
\begin{equation} \label{22}
\begin{array}{l}
\displaystyle{%
  dp = -2\upsilon n^2\frac{dV}{V} + \frac{5}{2}\,g\zeta(5/2)\frac{dT}{\Lambda^3}. %
}%
\end{array}
\end{equation}
In the result, we obtain
\begin{equation} \label{23}
\begin{array}{l}
\displaystyle{%
  \frac{C_V}{N}=\frac{3}{2}\frac{S}{N}, \qquad %
  \frac{C_p}{N}=\frac{3}{2}\frac{S}{N}\bigg[1+\frac{5}{6}\,g\zeta(5/2)\frac{T}{\upsilon n^2\Lambda^3}\bigg]. %
}%
\end{array}
\end{equation}
Taking into account the dependence of the total density on
temperature (\ref{16}), the isobaric heat capacity can be represented in the form %
\begin{equation} \label{24}
\begin{array}{l}
\displaystyle{%
  \frac{C_p}{N}=\frac{3}{2}\frac{S}{N} %
  \Bigg\{1+\frac{\sigma_0}{3\xi_p}\bigg(\frac{T}{T_P}\bigg)^{\!5/2}\frac{n_p^2}{n^2(0)} %
  \left[1-\bigg(1-\frac{n_p^2}{n^2(0)}\bigg)\!\bigg(\frac{T}{T_P}\bigg)^{\!5/2}\right]^{-1}\!\Bigg\}, %
}%
\end{array}
\end{equation}
where the quantity $\xi_p=\upsilon n_p/T_P$ is determined by the
dimensionless parameter $\eta$ (\ref{13})
\begin{equation} \label{25}
\begin{array}{l}
\displaystyle{%
  \xi_p=\frac{2}{5}\,\sigma_0\eta\bigg(\frac{T_P}{T_{P0}}\bigg)^{\!1\!/2}. %
}%
\end{array}
\end{equation}
Note that the following relation holds
\begin{equation} \label{26}
\begin{array}{l}
\displaystyle{%
  \frac{\sigma_0}{\xi_p}=\frac{5}{2}\bigg[\frac{n^2(0)}{n_p^2}-1\bigg]. %
}%
\end{array}
\end{equation}
The energy (\ref{10}) at $T<T_P$ is given by the formula
\begin{equation} \label{27}
\begin{array}{l}
\displaystyle{%
  \frac{E}{NT_P} = \xi_p\,\frac{n}{n_p} + \frac{3}{5}\,\sigma_0\bigg(\frac{T}{T_{P}}\bigg)^{\!5/2}\frac{n_p}{n}\,. %
}%
\end{array}
\end{equation}
At zero temperature, the energy per one particle $E(0)=\upsilon
n(0)=\sqrt{\upsilon p}$\, is determined by the interaction constant and pressure. %

\begin{figure}[b!]
\vspace{-00mm}  \hspace{0mm}
\includegraphics[width = 7.0cm]{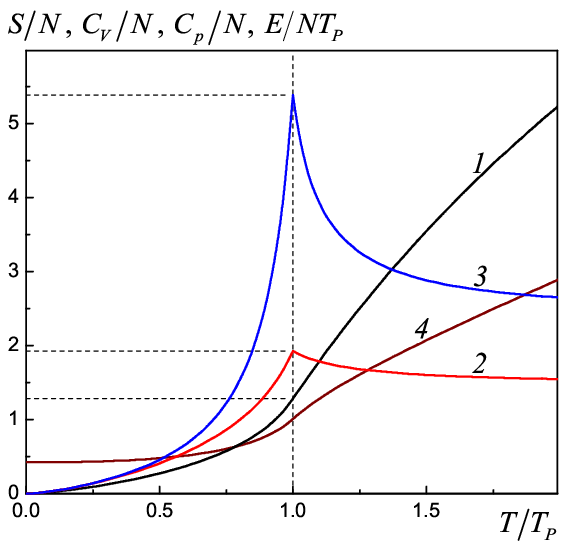} 
\vspace{-4mm} %
\caption{\label{fig04} 
Temperature dependences of the entropy, heat capacities and energy at $\eta=0.5$: %
{\it 1} -- $S/N=f(T/T_P)$,   {\it 2} -- $C_V/N=f(T/T_P)$, %
{\it 3} -- $C_p/N=f(T/T_P)$, {\it 4} -- $E/NT_P=f(T/T_P)$. %
At $T=T_P$: $S/N=1.28$, $C_V/N=1.93$, $C_p/N=5.39$. At $T=0$: $E/NT_P=0.42$. %
}%
\end{figure}

Near the condensation temperature the energy, entropy and heat
capacities depend linearly on temperature and at $T_P$ take on
finite values:
\begin{equation} \label{28}
\begin{array}{l}
\displaystyle{%
  \frac{S(T)}{N} = \sigma_0\Big(1+A_S^{(-)}\tau\Big), \qquad  %
  \frac{E(T)}{NT_P} =  \bigg(\xi_p+\frac{3}{5}\,\sigma_0\bigg)\Big(1+A_E^{(-)}\tau\Big), %
}\vspace{3mm}\\ %
\displaystyle{\hspace{0mm}%
  \frac{C_V(T)}{N} = \frac{3}{2}\,\sigma_0\Big(1+A_V^{(-)}\tau\Big), \,\,\,  %
  \frac{C_p(T)}{N} = \frac{3}{2}\,\sigma_0 \bigg(1+\frac{\sigma_0}{3\xi_p}\bigg)\Big(1+A_p^{(-)}\tau\Big), %
}%
\end{array}
\end{equation}
where $\tau\equiv\big(T-T_P\big)\big/T_P$, and the slope of lines is
determined by the coefficients
\begin{equation} \label{29}
\begin{array}{l}
\displaystyle{%
  A_S^{(-)}=A_V^{(-)}=\frac{1}{4}\left[1+5\,\bigg(\frac{n(0)}{n_p}\bigg)^{\!2}\right], \quad %
  A_E^{(-)}=\sigma_0\bigg(1+\frac{3}{10}\frac{\sigma_0}{\xi_p}\bigg)\bigg(\xi_p+\frac{3}{5}\,\sigma_0\bigg)^{\!-1}, %
}\vspace{3mm}\\ %
\displaystyle{\hspace{0mm}%
  A_p^{(-)}\equiv\frac{1}{4}\left[1+5\,\bigg(\frac{n(0)}{n_p}\bigg)^{\!2}\right] + %
  \frac{5}{6}\frac{\sigma_0}{\xi_p}\bigg(\frac{n(0)}{n_p}\bigg)^{\!2}\bigg(1+\frac{\sigma_0}{3\xi_p}\bigg)^{\!-1}.%
}%
\end{array}
\end{equation}
In contrast to the ideal gas model, where the isobaric heat capacity
in the phase with condensate goes to infinity \cite{YP2}, in this
case $C_p(T)$ turns out to be finite, so that all thermodynamic
relations are fulfilled \cite{YP1}. The temperature dependences of
the energy, entropy and heat capacities are shown in Figure \ref{fig04}. %

\section{Weakly non-ideal Bose gas with condensate at constant pressure
above the condensation temperature}\vspace{-0mm}
In the region $T>T_P$, the parameter $t$ is related to temperature by the relation %
\begin{equation} \label{30}
\begin{array}{l}
\displaystyle{%
  \xi_p\left[\bigg(\frac{T}{T_P}\bigg)^{\!3}\bigg(\frac{{\rm B}_{3/2}(t)}{\zeta(3/2)}\bigg)^{\!2}-1\right] + %
  \frac{2}{5}\,\sigma_0\left[\bigg(\frac{T}{T_P}\bigg)^{\!5/2}\,\frac{{\rm B}_{5/2}(t)}{\zeta(5/2)}-1\right]=0, %
}%
\end{array}
\end{equation}
which follows from the condition of constancy of pressure
(\ref{09}). Taking this relation into account, the temperature
dependence of density is calculated by the formula
\begin{equation} \label{31}
\begin{array}{l}
\displaystyle{%
  n=n_p\bigg(\frac{T}{T_P}\bigg)^{\!3/2}\frac{{\rm B}_{3/2}(t)}{\zeta(3/2)}. %
}%
\end{array}
\end{equation}
This dependence is shown in Figure \ref{fig03}.

The dependences of the entropy and isochoric heat capacity on the
parameter $t$ do not differ from the case of an ideal gas.  Taking
into account the relation $n\Lambda^3=g{\rm B}_{3/2}(t)$, we have
\begin{equation} \label{32}
\begin{array}{l}
\displaystyle{%
  \frac{S}{N}=\frac{5}{2}\left[\frac{{\rm B}_{5/2}(t)}{{\rm B}_{3/2}(t)}-t\right], %
}%
\end{array}
\end{equation} \vspace{-04mm}
\begin{equation} \label{33}
\begin{array}{l}
\displaystyle{%
  \frac{C_V}{N}=\frac{15}{4}\left[\frac{{\rm B}_{5/2}(t)}{{\rm B}_{3/2}(t)}-\frac{3}{5}\frac{{\rm B}_{3/2}(t)}{{\rm B}_{1/2}(t)}\right]. %
}%
\end{array}
\end{equation}
The formula for isobaric heat capacity, which was obtained in \cite{YP1}, %
\begin{equation} \label{34}
\begin{array}{l}
\displaystyle{%
  \frac{C_p}{N}=\frac{25}{4}\left[\frac{{\rm B}_{5/2}(t)}{{\rm B}_{3/2}(t)}-\frac{3}{5}\frac{{\rm B}_{3/2}(t)}{{\rm B}_{1/2}(t)}\right]\! %
  \frac{\displaystyle{ \left(\frac{{\rm B}_{5/2}(t)}{{\rm B}_{3/2}(t)}+\frac{6}{5}\,\xi\right) } }{\displaystyle{\left(\frac{{\rm B}_{3/2}(t)}{{\rm B}_{1/2}(t)}+2\,\xi\right)}}, %
}%
\end{array}
\end{equation}
includes the interaction constant, since $\xi\equiv\upsilon n/T$. In
the considered case of a fixed pressure, when the density depends on
temperature, we have
\begin{equation} \label{35}
\begin{array}{l}
\displaystyle{%
  \xi=\xi_p\bigg(\frac{T}{T_P}\bigg)^{\!1\!/2}\frac{{\rm B}_{3/2}(t)}{\zeta(3/2)}, %
}%
\end{array}
\end{equation}
where the quantity $\xi_p=(2/5)\,\sigma_0\eta\big(T_P/T_{P0}\big)^{1\!/2}$ (\ref{25}) %
is determined by the dimensionless parameter $\eta$. The energy at
$T>T_P$ has the form
\begin{equation} \label{36}
\begin{array}{l}
\displaystyle{%
  \frac{E}{NT_P}=\xi_p\bigg(\frac{T}{T_P}\bigg)^{\!3/2}\frac{{\rm B}_{3/2}(t)}{\zeta(3/2)} + %
  \frac{3}{2}\bigg(\frac{T}{T_P}\bigg)\frac{{\rm B}_{5/2}(t)}{{\rm B}_{3/2}(t)} . %
}%
\end{array}
\end{equation}
The temperature dependences of the energy, entropy and heat
capacities above $T_P$ are shown in Figure \ref{fig04}.

Near the condensation temperature at $t\rightarrow -0$ from (\ref{30}) we have %
\begin{equation} \label{37}
\begin{array}{l}
\displaystyle{%
  \sqrt{-t}=\frac{3\zeta(3/2)}{4\sqrt{\pi}}\bigg(1+\frac{\sigma_0}{3\xi_p}\bigg)\bigg(\frac{T-T_P}{T_P}\bigg). %
}%
\end{array}
\end{equation}
Here, in the same way as below $T_P$ (\ref{28}), the entropy, energy
and heat capacities depend linearly on temperature
\begin{equation} \label{38}
\begin{array}{l}
\displaystyle{%
  \frac{S(T)}{N} = \sigma_0\Big(1+A_S^{(+)}\tau\Big), \qquad  %
  \frac{E(T)}{NT_P} =  \bigg(\xi_p+\frac{3}{5}\,\sigma_0\bigg)\Big(1+A_E^{(+)}\tau\Big), %
}\vspace{3mm}\\ %
\displaystyle{\hspace{0mm}%
  \frac{C_V(T)}{N} = \frac{3}{2}\,\sigma_0\Big(1+A_V^{(+)}\tau\Big), \,\,\,  %
  \frac{C_p(T)}{N} = \frac{3}{2}\,\sigma_0\bigg(1+\frac{\sigma_0}{3\xi_p}\bigg)\Big(1+A_p^{(+)}\tau\Big), %
}%
\end{array}
\end{equation}
where
\begin{equation} \label{39}
\begin{array}{l}
\displaystyle{%
  A_V^{(+)}\equiv\frac{3}{2}\bigg(1-\frac{3}{4\pi\sigma_0}\big[\zeta(3/2)\big]^2\bigg)\!\bigg(1+\frac{\sigma_0}{3\xi_p}\bigg), %
}\vspace{3mm}\\ %
\displaystyle{\hspace{0mm}%
  A_p^{(+)}\equiv\frac{3}{2}\bigg(1+\frac{\sigma_0}{3\xi_p}\bigg)\!\bigg(1-\frac{3}{4\pi\sigma_0}\big[\zeta(3/2)\big]^2-\frac{\big[\zeta(3/2)\big]^2}{4\pi\xi_p}\bigg) + %
  \bigg(1+\frac{\sigma_0}{3\xi_p}\bigg)^{\!\!-1}\!\bigg(\frac{\sigma_0^2}{6\xi_p^2}-1\bigg)+\bigg(1+\frac{\sigma_0}{2\xi_p}\bigg), %
}%
\end{array}
\end{equation}
and $A_S^{(+)}=A_S^{(-)}$, $A_E^{(+)}=A_E^{(-)}$. Both heat
capacities are continuous at the transition temperature, but their derivatives experience a jump %
$\Delta\big(\partial C/\partial T\big)\equiv\big(\partial C/\partial T\big)_{T_P+0}-\big(\partial C/\partial T\big)_{T_P-0} $ %
during the transition from the high-temperature to the low-temperature phase: %
\begin{equation} \label{40}
\begin{array}{l}
\displaystyle{%
  \Delta\bigg(\frac{\partial C_V}{\partial T}\bigg)=\frac{3\sigma_0}{2}\frac{N}{T_P}\Big(A_V^{(+)}-A_V^{(-)}\Big),  %
}%
\end{array}
\end{equation}\vspace{-3mm}
\begin{equation} \label{41}
\begin{array}{l}
\displaystyle{%
  \Delta\bigg(\frac{\partial C_p}{\partial T}\bigg)=\frac{3\sigma_0}{2}\bigg(1+\frac{\sigma_0}{3\xi_p}\bigg)\frac{N}{T_P}\Big(A_p^{(+)}-A_p^{(-)}\Big).  %
}%
\end{array}
\end{equation}
As in the case of the transition at constant density \cite{YP1}, the
energy and entropy and their derivatives are continuous at the
transition temperature.

\section{Conclusion}\vspace{-02mm}
In the model of a weakly non-ideal Bose gas \cite{YP1}, the
Bose-Einstein condensation at constant pressure is considered. The
condensation temperature, depending on the pressure and the
interaction constant, is found. The temperature dependences of the
total density and the density of Bose-Einstein condensate, the
energy, entropy, isochoric and isobaric heat capacities are
obtained. At the condensation temperature, the derivatives of the
heat capacities experience a jump, while the energy, entropy and
their derivatives are continuous.

The author thanks A.A.\,Soroka for help in preparing the article.


\vspace{05mm}
\begin{thebibliography}{99}
\bibitem{YP1}
Yu.M.\,Poluektov, {\it A simple model of Bose-Einstein condensation
of interacting particles}, J. Low Temp. Phys. {\bf 186}, \linebreak
347\,--\,362 (2017). doi:10.1007/s10909-016-1715-5; arXiv:1602.02746v1\,[cond-mat.stat-mech] %
\bibitem{YP2}
Yu.M.\,Poluektov, {\it Isobaric heat capacity of an ideal Bose gas},
Russ. Phys. J. {\bf 44}(6), 627\,--\,630 (2001). doi:10.1023/A:1012599929812 %
\end{thebibliography}
\end{document}